# PREFER: An Ontology for the PREcision FERmentation Community

Txell Amigó[1], Shawn Zheng Kai Tan[2], Angel Luu Phanthanourak[1], Sebastian Schulz[1], Pasquale D. Colaianni[1], Dominik M. Maszczyk[1], Ester Milesi[1], Ivan Schlembach[1], Mykhaylo Semenov Petrov[1], Marta Reventós Montané[1], Lars K. Nielsen[1,3], Jochen Förster[1], Bernhard Ø. Palsson[1,4], *Suresh Sudarsan[1, 5]\*, and Alberto Santos[1]\**

1 The Novo Nordisk Foundation Center for Biosustainability, Technical University of Denmark, Kgs. Lyngby, Denmark †
2 SignaMind, Singapore
3 Australian Institute for Bioengineering and Nanotechnology, The University of Queensland, Brisbane, Queensland, Australia.
4 The Department of Bioengineering, University of California, San Diego, USA
5 Nexxar ApS, Lynge, Denmark

*Correspondence to: albsad@dtu.dk; suresh.sudarshan@gmail.com
†Present address: The Novo Nordisk Foundation Biotechnology Research Institute for the Green Transition (BRIGHT), Technical University of Denmark, Kgs. Lyngby, Denmark

## Abstract

Precision fermentation relies on microbial cell factories to produce sustainable food, pharmaceuticals, chemicals, and biofuels. Specialized laboratories such as biofoundries are advancing these processes using high-throughput bioreactor platforms, which generate vast datasets. However, the lack of community standards limits data accessibility and interoperability, preventing integration across platforms. In order to address this, we introduce PREFER, an open-source ontology designed to establish a unified standard for bioprocess data. Built in alignment with the widely adopted Basic Formal Ontology (BFO) and connecting with several other community ontologies, PREFER ensures consistency and cross-domain compatibility and covers the whole precision fermentation process. Integrating PREFER into high-throughput bioprocess development workflows enables structured metadata that supports automated cross-platform execution and high-fidelity data capture. Furthermore, PREFER's standardization has the potential to bridge disparate data silos, generating machine-actionable datasets critical for training predictive, robust machine learning models in synthetic biology. This work provides the foundation for scalable, interoperable bioprocess systems and supports the transition toward more data-driven bioproduction.

## Introduction

### Background

Precision fermentation is a core technology in the emerging bio-based economy, enabling the sustainable production of chemicals and materials through microorganisms engineered for fermentation processes [1]. Applications of precision fermentation now include energy (e.g. biofuels from companies such as LanzaTech)[2], food (e.g. plant-based meat substitutes from companies such as Impossible Foods)[3], agriculture (e.g. biofertilizers produced by companies such as Pivot Bio)[4], and many more. Despite these advances, significant challenges remain in scaling processes from laboratory to industrial systems [5]. This is mainly due to the lack of accessibility of bioprocess data and metadata, collected before, during and after fermentation, which are critical for identifying scale, strain and condition-dependent parameters. Bridging this gap in bioprocess development and scalability is essential to reduce time and cost in biomanufacturing [2].

Biofoundries, positioned at the interface between academia and industry, play a critical role in addressing the challenges in bioprocess scale-up [6]. In these research institutions, the implementation of high-throughput bioreactor systems/platforms allows parallel experiments, enabling rapid testing and generation of large amounts of data. This screening approach allows researchers to evaluate synthetic engineered biocatalysts at a smaller scale, identifying high performing cell factories for target products. Only the best-performing strains progress to a larger-scale testing. This tiered screening strategy substantially reduces the experimental burden and resource requirements associated with exploring larger bioreactor volumes. All the high-dimensional data generated within these biofoundry infrastructures represent a high-value asset beyond the scope of any individual experiment. Currently, much of this value remains locked in siloed datasets. To fully benefit, disparate experimental streams would need to be harmonized into a unified analytical framework. However, integrating data across diverse bioreactor platforms, operational modes, measurement techniques and calculation types remains a critical challenge [6,7]. Proprietary software and heterogeneous data outputs provided by different platforms further complicate standardizing resources. This lack of standardized data practices within and across institutions has limited the development of tools for analyzing and visualizing bioprocess data, the ability to exploit multi-omics data, and the application of Artificial Intelligence (AI) to optimize precision fermentation processes effectively [8]. Advancing bio-based production therefore depends heavily on improved bioprocess data management and, above all, data interoperability.

To this end, the FAIR principles (Findability, Accessibility, Interoperability, and Reusability)[9] have become essential guidelines for data-driven biomanufacturing [2]. Initiatives such as the Global Biofoundry Alliance (GBA) [10], online software tools like EDD (https://edd-docs.jbei.org/), along with infrastructures like The National Institute of Standards and Technology (NIST Biofoundry) (https://www.nist.gov) and the Industrial Biotechnology Innovation and Synthetic Biology Accelerator (IBISBA), are actively promoting data standardization, open protocols, and collaborative practices. Despite these efforts, widely adopted open-source frameworks and community-driven platforms that address computational and data-management challenges for precision fermentation are scarce. This lack of accessible tools limits the broader adoption of FAIR principles across academic and industrial settings.

Ontologies, and their related data management systems, are a powerful approach for realizing the FAIR principles [11]; they improve findability by establishing a semantic framework that supports complex querying, and they provide controlled vocabularies that enhance data interoperability and promote clear data understanding, which in turn facilitates data reusability [12]. Furthermore, data aligned to ontologies can be directly used in AI applications through automated reasoning and integration into machine learning pipelines [13].

While there are a few existing legacy ontologies in the fermentation domain, their scope is limited to traditional fermentation [14] and fermented food applications [15]. Consequently, these ontologies are insufficient and fail to capture the technological specificity and complexity inherent to precision fermentation processes. To bridge this gap and provide the community with necessary standardisation capabilities, we introduce the PREcision FERmentation (PREFER) ontology. PREFER is a comprehensive semantic framework designed to integrate high-throughput bioprocess data, covering operational, environmental and process parameters across different scales of a precision fermentation process, to accelerate the development and scaling of biosustainable production processes.

### Rationale

Sustainability is a primary objective in industrial biotechnology; achieving a viable bioeconomy requires aligning environmental sustainability with economic feasibility. This efficiency aim is increasingly targeted through the integration of artificial intelligence (AI). Despite the advancement of predictive tools in strain development [16–18], scaling engineered strains into efficient production organisms and full-scale factories continues to pose major challenges. This is largely due to our limited ability to extract actionable information for improving biomanufacturing from the large amount of data being generated [19,20].

At the heart of this challenge lies data. If biofoundries and other specialized laboratories are to collaborate effectively, their data must adhere to FAIR principles so that information generated across diverse experiments and platforms can be seamlessly integrated, interpreted, and leveraged for innovation. With this in mind, we developed PREFER to introduce semantic structure to bioprocess data, enhancing its interoperability, and ensuring that the data is AI-ready. By enabling such integration, PREFER helps make bioprocesses more competitive and supports manufacturing with living cells as a truly scalable biosolution for a sustainable bioeconomy.

PREFER was built in close collaboration with domain experts to ensure accurate representation of bioprocess concepts and workflows. Our initial objective was to create a unified environment where datasets from different experiments could be collected, visualized, and integratively computed. Such a system would require a flexible structure suited for analytics of interrelated datasets (data and metadata) originating from bioreactor instrumentation, laboratory information management systems (LIMS), and manufacturing execution systems (MES) such as PAS-X (https://www.koerber-pharma.com/en/solutions/software/werum-pas-x-savvy).

During the development of a data model for such a system, it became clear that a controlled vocabulary was missing. We therefore created a terminology comprising well-defined concepts relevant to bioprocessing. As patterns in the conceptual structure emerged, this terminology evolved into a taxonomy, classifying terms into meaningful hierarchies. Recognizing the need for a more expressive, machine-understandable framework that aligns with Industry 4.0 principles and supports cyber–physical systems [21–23], we extended this work beyond a terminology and taxonomy to develop a full ontology, PREFER, capable of standardizing bioprocess knowledge in a computationally interpretable form.

## Results

The ontology models the comprehensive data flow of a bioproduction process, with inputs such as strain, media components, and instruments, and outputs such as products, by-products and their associated measured and computed variables {Figure 1}.

The biomanufacturing process begins with a bioreactor unit containing a fermentation culture and can be explored by zooming into the cell factory process. At this level, a microbial strain consumes substrates and converts them into products and by-products, while simultaneously generating biomass.

The fermentation run unfolds over time, progressing from the initial setup of bioreactor operating setpoints and starting conditions to the continuous acquisition of process data. Throughout this dynamic process, data are collected through online measurements (e.g. pH, temperature, pressure, volume) and offline sampling (e.g. biomass sample). These heterogeneous data streams encompassing measured and calculated variables, and omics data, form the foundation for downstream analysis. The formal and interoperable representation of these data is achieved through the PREFER ontology, which is detailed in the subsequent section.

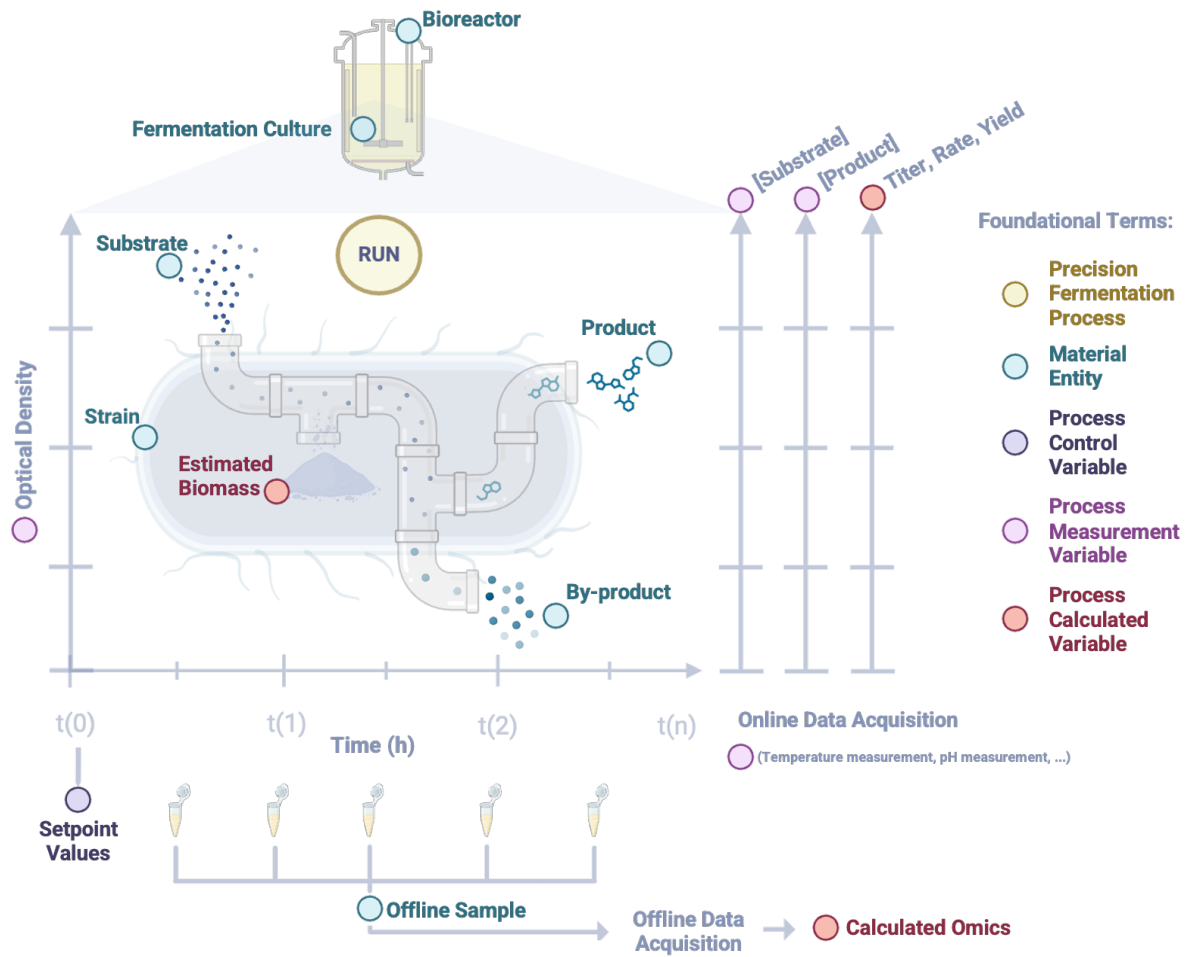

**Figure 1. A journey of a single bioreactor run: exploring a precision fermentation process.**
Overview of a fermentation run illustrating how experimental data and metadata are mapped, using colored nodes to highlight the core concepts and associated semantics of the PREFER ontology (foundational terms shown on the right). The figure is organized from top to bottom, moving from the bioreactor to a zoom into the fermentation culture and cell factory process, and finally showing a timeline at the bottom depicting data collection and analysis over the course of the run.

### Ontology Design

The PREFER ontology is constructed as a Basic Formal Ontology (BFO)-conformant ontology to ensure logical coherence and interoperability across domains. To assure high standards, PREFER was also developed based on the Open Biological and Biomedical Ontologies Foundry (OBO Foundry) [24] and FAIR principles [9]. It comprises eight foundational top-level terms, of which five are core terms, the central conceptual categories needed to describe the precision fermentation process, and three are supporting terms providing context, structure and semantic consistency to the core entities {Figure 2}. These foundational terms are either direct BFO terms or subclasses of BFO terms. This structured approach aims to provide a rigorous semantic foundation that supports interoperability, extensibility, and clear representation of process-centric knowledge in precision fermentation. PREFER being OBO and BFO conformant also means that it can be easily integrated with other ontologies to cover more processes related to the precision fermentation field (e.g. bio-ontologies like the Gene

Ontology (GO) [25], Phenotype And Trait Ontology (PATO) [26]), and even financial ontologies like the Financial Industry Business Ontology (FIBO) [27], which is partially BFO conformant, to model financial concepts like expected revenue. This is particularly relevant when linking bioprocess data to systematic economic evaluation methods such as Techno-economic Assessment (TEA)[28], a biotechnology framework that integrates technical performance with economic feasibility to assess process efficiency, costs, and market potential.

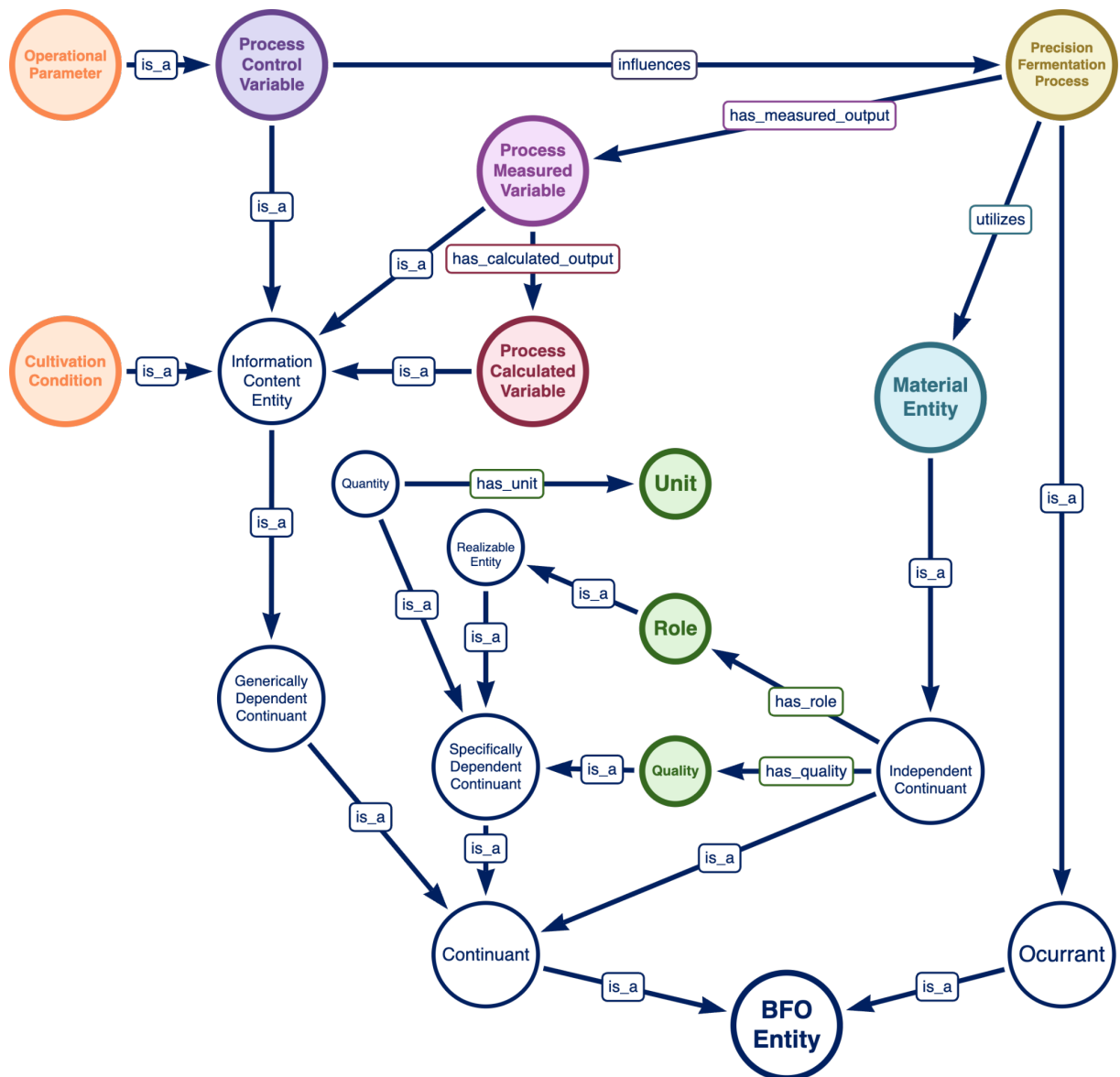

**Figure 2. Overview of the foundational terms of the PREFER ontology and their alignment with the Basic Formal Ontology (BFO).**
Core PREFER foundational terms are distinguished from BFO terms by colored backgrounds (yellow, purple, pink, light blue, and red), supporting terms are highlighted in green, and polyhierarchical terms are shown in orange.

Within this framework, the core top-level concepts are 1) Process Control Variable, which serves to represent inputs to the process; 2) Precision Fermentation Process itself, which models processes and events; 3) Process Measured Variable, representing measurements taken from the process; 4) Process Calculated Variable, which are derived or calculated from measured variables; 5) Material Entity, which consists of physical entities utilized within the processes. We envision this covering data from the whole precision fermentation process. For example, an experiment (a type of Precision Fermentation Process), utilizes Material Entities such as a bioreactor, fermentation medium, and fermentation culture. The fermentation requires certain parameters such as temperature and pH setpoints to be controlled, which are modelled as Process Control Variables. Information such as temperature and pH readings are recorded during the process which are modelled as Process Measured Variables. Information might then have to be secondarily calculated (e.g. rates, yields, and titers), these are represented as Process Calculated Variables.

Figure 2 shows that 'process measured variable' and 'process control variable' are modeled as a polyhierarchy with parent terms "cultivation condition" and "operational parameter". This polyhierarchy, where a single item can be a member of more than one category (multiple "parents" or broader terms), allows simultaneous separation between control and measurement variables. This avoids an artificial separation of related terms and supports more flexible querying and semantic reasoning across process inputs and outcomes. For instance, pH related variables in Process Control Variables (e.g. pH setpoint) and Process Measured Variables (e.g. pH measurement) can be classified under a Cultivation Condition (e.g. alkaline cultivation condition). Operational parameters are a subset of Process Control Variables that represent a set of configurable settings used to control how cultivation conditions are regulated in a bioreactor or fermentation system during the process.

The three supporting parent classes are role, quality, and unit. Role and quality are used to describe material entities (e.g. taxon as a quality to describe the microorganism [29] or seed culture versus production culture as a role of the fermentation culture during a fermentation process), and units are used to standardize and quantify physical quantities.

Variables (Process Control Variable, Process Measured Variable, and Process Calculated Variable) are modelled as information content entity from the Information Artifact Ontology (IAO), which is classified as generically dependent continuant in BFO. The decision to model Process Control Variable as an information entity instead of a realizable entity was done for simplicity. Realizable entities are dependent on a physical bearer and realized through processes, which present challenges in situations where control is abstract, composite, or distributed. For example, some control feedback loops can be directly adjusted by the user (e.g. a Proportional-Integral-Derivative (PID) loop for pH control), while others are proprietary or not directly accessible (e.g. temperature control). In many cases, the control specification does not directly correspond to a physical component. For example, there is no physical dial for temperature 37 °C, instead temperature is regulated by controlling heating and cooling water flow through software-defined or algorithmic mechanisms, whereby the tangible instruments would be a complex set of valves, pumps and heating elements. By taking Process Control Variable as a specification or directive, the model decouples the abstract input from its complex physical realization. This distinction allows complex control parameters to be simply represented as informational inputs, avoiding the necessity of modeling the whole process of control.

**Community Development**

The development of the PREFER ontology was intentionally designed with community contribution in mind to foster collaboration and continuous improvement. Our approach was to first establish a terminology with the logical foundations of an ontology, and then iteratively expand logical axioms and definitions through community-driven use cases, explicitly inviting collaboration. This will allow immediate adoption of ontology IDs as a controlled vocabulary, which in itself will aid interoperability of data, enable FAIR@source (or "born FAIR") [30], and reduce the need for harmonization down the road.

By leveraging the Ontology Development Kit (ODK) [31], PREFER ensures that its GitHub repository follows standardized structures and conventions, lowering the barrier for new contributors by enabling anyone familiar with the ODK to easily understand the project layout and begin contributing effectively. Additionally, hosting the ontology on GitHub allows community members and subject matter experts who may not have ontology editing experience to engage by opening tickets or issues, facilitating feedback and requests for changes. The PREFER ontology GitHub repository also features an on-demand action that uses ROBOT [32] to generate human readable diffs on pull requests, simplifying the review process and reducing technical hurdles for contributors. Lastly, as PREFER uses widely accepted standards and tooling, contributors can access existing guides and tutorials like those developed by and for the OBO academy [33], for example, the obook (https://oboacademy.github.io/obook/). Our aim for this was to develop an open, transparent, and well-supported development environment, ensuring that PREFER can grow as a living resource tailored by its community of experts and users.

## Discussion

In this manuscript, we describe the PREFER ontology, the reasons it was developed, and the rationale behind its design and construction. Beyond that, we suggest how the ontology can be used and how the community can not only utilize the resource but also contribute back and co-develop this as a solution to harmonizing the disparate sources of data.

Precision fermentation is advancing rapidly as a transformative technology to enable high quality sustainable production [3], but the field faces significant lack of standardized data frameworks hampering the integration of experimental results across different scales, bioreactor types, and software platforms, limiting the ability to accelerate process development cycle time. The PREFER ontology is designed to solve these challenges by providing a structured, standardized terminology that enables interoperability across precision fermentation data sources and experimental conditions, allowing for development of standardized visualization and analytical tools for precision fermentation experiments across different cultivation platforms and bioreactor scales.

Beyond integration of data, as PREFER is built as an ontology rather than a controlled vocabulary, thesaurus, or taxonomy, it provides a formal, machine-interpretable representation of the precision fermentation domain that directly enables symbolic AI applications [34]. This formal structure supports advanced reasoning tasks such as consistency checking and inference of implicit relationships within the data. Aligning data to PREFER also facilitates the creation of high-quality semantic knowledge graphs, which serve as a foundation for neuro-symbolic AI approaches [35]. These approaches combine the strengths of reasoning with AI models, paving the way toward more explainable and trustworthy AI systems, an essential feature for advancing precision fermentation technologies where transparency and interpretability are critical for adoption and regulatory compliance.

More specifically, PREFER is a crucial component towards building simulators based on real bioprocess data for scale-down and scale-up contexts, an important area of ongoing development in precision fermentation [36,37]. Data aligned to PREFER can be more optimally utilized to optimize process parameters and to better guide the Design of Experiments (DoE) in biomanufacturing.

We further view PREFER as a key enabler for addressing scale-up challenges during the transition from screening to biomanufacturing, bridging between strain engineering and process engineering. For example, by covering the entire bioprocess window, PREFER can support a deeper understanding of strain physiological requirements and the implications these have for metabolic pathway design.

At the most fundamental level, PREFER serves as a terminology layer for precision fermentation, providing a controlled vocabulary (CV) for key bioprocess variables. In this role, it supports interoperability and consistent data management across information systems. For example, through the implementation of PREFER as a CV, certain scale-up questions can be answered via standard retrieval queries, such as:

> *"Give me all rates for runs conducted in 250 mL and 5L bioreactors for tyrosine products."*

Here, PREFER terms standardize the vocabulary and units for variables such as calculated rates, reactor volumes, and product identifiers, enabling access to comparable data across experiments and different scales.

Beyond this terminological and data management function, PREFER is also an ontology with an explicit semantic structure and reasoning model that enables more advanced, inference-driven use cases. In this capacity, it supports symbolic AI workflows, where formal logical axioms are used to derive new knowledge from existing data. For instance, one can ask:

> *"For this target product, is there any related product that shares the same upstream metabolic pathway, such that we can infer from historical bioprocess data whether oxygen-limited or oxygen-unlimited conditions are likely more advantageous for the production of this type of target product?"*

In this case, the ontology not only enables comparison of fermentation runs across different scales but also supports extracting implicit knowledge through a reasoner. By linking to other ontologies such as ChEBI, GO, etc., PREFER bridges multiple domains of knowledge, such as fermentation process, product identity, and gene expression, allowing logical inference about how cultivation conditions (e.g. oxygen availability) influence gene expression levels and metabolic pathways associated with a given target product. These forms of inferencing enable scientists to observe correlations between oxygen levels and product formation and potentially unlock additional metabolic interventions or process strategies that may further enhance product yields.

PREFER is designed not only to link products to gene-related knowledge domains (e.g. genes, metabolic processes, and pathways), but also to integrate additional process-related domains and their associated logic, including downstream processing and sustainability-related considerations. These may include, for example, the energy costs associated with specific operational parameters (e.g. stirring required to maintain dissolved oxygen setpoints) or the costs and environmental impacts of compounds used in media formulations.

Taken together, these two layers, (i) PREFER as a terminology and data management backbone, and (ii) PREFER as a reasoning-capable ontology, are intended to model the logic of a fermentation process as a generalizable semantic structure that enriches precision fermentation data with explicit meaning. This framework supports the consolidation and systematic reuse of the knowledge required to advance bioprocesses.

Because we envision the PREFER ontology as a shared resource for the precision fermentation community, this manuscript serves as a call to co-develop PREFER. We aim to develop the ontology's logical axioms and definitions based on use cases from the community to cover the diverse needs across fermentation scales, modalities, and applications. For this reason, we made PREFER open-source and opened it for contributions to our terminology, definitions, and other annotations through adding issues to our GitHub repository. This model will allow the sustainable development of PREFER through shared projects and/or direct contributions to the repository, which is helping to maintain bio-ontologies in general [12].

Looking into the future, PREFER can be further developed for use as an integration and logic layer that enables operational AI in bioprocessing, a crucial step towards developing digital twins for precision fermentation processes [22]. Having data aligned and interoperable will aid in developing machine learning models that can greatly aid in the development of digital twins, while having ontologies that model and hence integrate data for disparate sources and sensors is a crucial step towards adaptive digital twins [38]. These digital twins can then aid in yield optimization, predictive maintenance, quality control automation, and much more.

**Potential Application and Future Plans**
Following the development and release of PREFER, our next steps would be to link biofoundry data up the ontology, and from there develop semantic knowledge graphs around them. The goal is to enable interoperability through data standardization, allowing harmonized data to be reused with community analytical and AI-based predictive tools for faster and more effective evaluation of fermentation processes with candidate production strains. To achieve this, we propose a federated architecture and suggest ways in which the biofoundry and bioprocess labs can start adapting PREFER in this section.

One potential application of PREFER is the inference of missing metadata through semantic relationships defined in the ontology. For example, a laboratory may perform fermentations using only a single operational mode,batch, fed-batch or continuous, but may not explicitly record this information in the dataset. By aligning the data to the PREFER ontology, the operational mode can be inferred from the feeding behaviour of the process, such as the presence of feed rate control variables. Based on this information, the process can be classified as operating in fed-batch or continuous mode.
This would enable the classification and comparison of different fermentation modes without requiring users to explicitly capture this metadata. As a result, the ontology would support meaningful comparison of experiments, even across different operational modes and bioreactor scales and simplify metadata collection.

To link experimental data to PREFER in a way that accommodates the diversity of biofoundry workflows, we envision a layered architecture where PREFER serves as the domain ontology defining and classifying core bioprocess concepts. Data can be integrated either by directly using PREFER URIs (e.g. using PREFER:0000203 instead of a text string "production culture") or indirectly using an application ontology that acts as a bridge between domain-specific implementations and PREFER. These potential application ontologies will enable the use of local data models and terminologies, connecting them to PREFER and other datasets through logical mappings. For example, while most of the microorganisms' growth are classified based on their optimal temperature and pH, and if a cell factory process does not alter their setpoints, but registers them as "default condition setpoint", an application ontology can convert them to:

'default condition setpoint' equivalentTo
  'process control variable'
  and ('has setpoint' value 'default temperature setpoint')
  and ('has setpoint' value 'default pH setpoint')

'default temperature setpoint' equivalentTo
  'temperature setpoint control variable'
  and ('has unit' some 'degree Celsius')
  and ('has value' exactly 37)

'default pH setpoint' equivalentTo
  ('pH setpoint control variable'
  and ('has unit' some 'pH based unit')
  and ('has value' exactly 7))

This allows the laboratory to continue registering their metadata as default condition but still interoperate with data from labs that register their pH setpoints and temperature setpoints separately. By expanding collapsed metadata into explicit control variables and retrieving implicit defaults when critical parameters are missing, the application ontology enables semantic comparability across experiments and supports meaningful cross-laboratory data integration.

This architecture enables federation of governance, distributing governance responsibilities across institutions, while allowing each to retain control over its own application ontologies, and still supporting global interoperability. However, realizing this approach requires institutes to have capabilities to utilize ontologies, and even manipulate and create ontologies with all the nuances of formal logic if they were to choose the application ontology route. Modelling languages such as LinkML [39] are well-suited for such tasks and natively have support for ontologies. Further work to develop such artefacts is required for optimal use of PREFER.

Our suggested approach to using PREFER would be:
 1) Identifying core process parameters and variables within your system.

2) Mapping these to PREFER (either directly or through an application ontology).
3) Utilising the URIs (either PREFER or application ontology IDs) in your database.
4) Virtualising or materialising, dependent on needs, into a semantic knowledge graph that has PREFER as the underlying ontology.

A simple and effective starting point would be to begin with steps 1-2, identifying and harmonizing bioprocess data to PREFER, thereby enabling interoperability and facilitating the implementation of analytical and predictive tools.

Overall, we believe our proposed approach is a practical way to allow the flexibility needed in biofoundries and bioprocess labs while still ensuring semantic consistency, interoperability, and scalability across bioprocessing data sources.

**Methods**

**Building PREFER**

The PREFER ontology is developed using the Ontology Development Kit (ODK) [31]. Following OBO Foundry [24] principles, it utilizes terms for existing ontologies where sensible. It currently imports terms from the Information Artifact Ontology (IAO) [40] which includes BFO [41], Cell Ontology (CL) [42,43], OBO Metadata Ontology (OMO), OBO Relation Ontology (RO) [44], Chemical Entities of Biological Interest Ontology (ChEBI) [45], Bioassay Ontology (BAO) [46], PATO [26], and GO [25,47].
PREFER aims to be compliant with OBO Foundry standards and is in the process of seeking admission into the OBO Foundry. Manual edits were done using Protege [48]. Final products were materialised through ROBOT [32] implementation in ODK using an ELK reasoner [49].

**Ontology Content Metrics**
To obtain statistics, the final product was loaded into local Ontotext GraphDB triple store (https://www.ontotext.com/products/graphdb/) and using the SPARQL query found in https://oboacademy.github.io/obook/reference/sparql-reference/?h=sparq#count-class-by-prefixes and modifying for specific needs.
As of the v2025-12-22 version, PREFER contains 174 classes and 22 object properties in the prefer namespace, all other terms are imported from external ontologies.

**Data Availability**

PREFER can be accessed through OLS Ontology Lookup Service [50,51] (https://www.ebi.ac.uk/ols4/ontologies/prefer) and Bioportal [52] (https://bioportal.bioontology.org/ontologies/PREFER).

PREFER available for download from its GitHub release in two forms (full and base) and formats (OWL, OBO and OBOgraphs JSON). The main release can be downloaded from:

https://github.com/Multiomics-Analytics-Group/prefer_ontology/releases/latest/download/prefer.owl

All other release products, where full is the same as the main release, and base is a product with links to external ontologies recorded as bare IRIs, designed to be combined with other base ontologies to make integrated products and for use in generating import modules, can be downloaded from:

https://github.com/Multiomics-Analytics-Group/prefer_ontology/releases/latest/download/prefer-{full/base}}.{owl/obo/json}

Programmatic access is available via the Ontology Lookup Service API (https://www.ebi.ac.uk/ols4/api/ontologies/prefer/), the Bioportal API (see http://data.bioontology.org/documentation)

## Code Availability

All code presented can be found in the PREFER github repository (https://github.com/Multiomics-Analytics-Group/prefer_ontology)

## Funding

This project was supported by the Novo Nordisk Foundation grant NNF20CC0035580.

## Author Contributions

A.S and S.Sudarsan conceived the project. T.A., S.Z.K.T. and A.S. designed and implemented PREFER. M.R.M., M.S.P., I.S., E.M., S.Schulz and S.Sudarsan provided expertise on the fermentation process and associated variables. A.L.P., P.D.C and D.M.M provided support regarding the informatics infrastructure. T.A., S.Z.K.T. and A.S. wrote and edited the manuscript and prepared the figures. A.S. directed the project. All authors reviewed and approved the final manuscript.

## Competing Interests

The authors declare no competing interests.

## Acknowledgements

We thank all members of the Multi-omics Network Analytics research group at the Novo Nordisk Foundation Center for Biosustainability for their help providing feedback. Similarly, we thank the Integrated Fermentation Platform and Pre Pilot Plant research and translational teams at the Novo Nordisk Foundation Center for Biosustainability, for their support and technical explanations, likewise the Quantitative Modelling of Cell Metabolism research team lead by Lars Keld Nielsen for their guidance and detail clarifications. We also thank Dr. Stefano Donati and his research group at DTU Biosustain for his support and insights on the way they perform fermentations. We would also like to acknowledge Lea M. Sommer for her input on the potential of PREFER. Figure 1 was created with BioRender.com (BioRender.com/bqcybqp).